# An Investigation of Wavelet Packet Transform for Spectrum Estimation

D.D. Ariananda, M.K. Lakshmanan, H. Nikookar
International Research Center for Telecommunications and Radar (IRCTR)
Department of Electrical Engineering, Mathematics and Computer Science
Delft University of Technology, Mekelweg 4, 2628 CD Delft, The Netherlands
m.k.lakshmanan@tudelft.nl, h.nikookar@tudelft.nl

*Abstract-* In this article, we investigate the application of wavelet packet transform as a novel spectrum sensing approach. The main attraction for wavelet packets is the tradeoffs they offer in terms of satisfying various performance metrics such as frequency resolution, variance of the estimated power spectral density (PSD) and complexity. The results of the experiments show that the wavelet based approach offers great flexibility, reconfigure ability and adaptability apart from its performances which are comparable and at times even better than Fourier based estimates.

## I. INTRODUCTION

The problem of spectrum estimation is in evaluating and identifying the distribution in frequency of the power of a signal amidst harsh and noisy environments. Metrics like speed and accuracy of measurement as well as complexity of method are important considerations when choosing the right spectrum analysis tool. Accuracy corresponds to frequency resolution, bias or leakage and variance of the estimated PSD. Since wireless nodes are envisioned to operate with small size and power, complexity of implementation is a key consideration.

Wavelet packet transform is the latest addition to the rich arsenal of communication system tool box. They have been known for their promising performance when they are used as a basis for new multi-carrier communication technique [1]. Apart from this, wavelets are also recognized as powerful mathematical tools to analyze local structure of frequency spectrum to identify singularities and edges. The main attraction for wavelets in this application is in their ability to analyze singularities and irregular structures and the tradeoffs they provide in terms of handling metrics such as frequency resolution of the spectra, variance of estimated power spectra and complexity.

In this paper, we present a spectrum estimation approach based on wavelet packet transform. To this end, in Section II, we first state traditional spectrum estimation techniques, such as Periodogram, Blackman-Tukey, Welch and Multi taper estimation, as filter bank analysis problem. Then in Section III we present the Wavelet Packet based spectrum estimation, implemented using paraunitary filters, as advancement to the existing approaches. To corroborate the theoretical analysis a few results are provided in Section IV. The conclusions are drawn in section V.

## II. FILTER BANK PARADIGM AND SPECTRUM ESTIMATION

### A. Periodogram, Blackman-Tukey and Welch

Spectrum estimation is about finding the power spectrum density (PSD) of a finite sample set $\{x(n-k), k = 0, 1, \ldots, N-1\}$ for frequency $|\omega| \leq \pi$. The classical approach to spectrum estimation is to use Fourier transforms to obtain a Periodogram, given as [4]:

$$\hat{PSD}_{PERIOD}(f) = \frac{1}{N}\left|\sum_{n=0}^{N-1} x(n)e^{(-j2\pi fn)}\right|^2 \quad (1)$$

For any given frequency $f_i$, (1) can be written as:

$$\hat{PSD}_{PERIOD}(f_i) = \left|\sum_{k=0}^{N-1} h_i(k)x(n-k)\right|^2 \quad (2)$$

where $\quad h_i(k) = w(k)e^{j2\pi f_i k} \quad (3)$

and $w(k)$ is a window function. If $w(k)$ is taken to be a prototype FIR lowpass filter, then $h_i(k)$ s will constitute a bank of bandpass filters centered at frequencies $f_i$s. The periodogram estimate may then be viewed as the output of several filters banks, constructed by modulating a single prototype filter $w(k)$, with each point in the PSD estimate corresponding to a filter's output. At its simplest the periodogram applies a rectangular window i.e. $w(k) = 1/\sqrt{N}$. Such a window function is not desirable because a rectangular window in time is a sinc function in frequency thereby resulting in a high side lobe and large leakages in the estimates. This problem can be alleviated by replacing the rectangular window with a window function with a taper that smoothly decays on both sides to obtain a prototype filter with much smaller side lobes. A few popular windows are Hamming, Hann and Blackman [2]. Another problem with the periodogram is that the estimates of the PSD are coarse with low precision and large variance which *does not* improve with more data. The only way to improve the variance of the periodogram is to average the PSD coefficients. This can be done in two ways: one by computing several (shorter) periodograms and use these to compute averages of each PSD coefficient. Or by applying a window function to the data first. This is equivalent to replacing each periodogram coefficient by a weighted average of coefficients. This is what happens in the Blackman-Tukey method [2]. And of course, the two methods can be combined, so that one computes an average of several windowed periodograms. This is the Welch method [2].

### B. Multi Taper Spectrum Estimation

The Multi Taper Spectrum Estimator (MTSE), proposed by Thomson [3], uses multiple orthogonal prototype filters to improve the variance and reduce the sidelobe and leakage. The process is initiated by collecting the last $M$ received samples in a vector $\mathbf{x} = [x(n)\ x(n-1)\ \ldots\ x(n-M+1)]^T$ and representing it as a set of orthogonal slepian base vectors [4]:

$$\mathbf{x}(n) \approx \sum_{k=0}^{K-1} \kappa_k(f_i)\, \mathbf{D}\, \mathbf{q_k} \quad (4)$$

In (4), $\kappa_k(f_i)$ is the expansion coefficients, $K$ is the total number of orthogonal prototype filters, $\mathbf{q_k}$ is the set of orthogonal slepian basis vectors (prolate spherical sequences) derived using a minimax algorithm and $\mathbf{D}$ is a diagonal matrix with the diagonal elements of 1, exp(j2π$f_i$), ...., exp(j2π(M-1)$f_i$). $\kappa_k(f_i)$ is given as:

$$\kappa_k(f_i) = (\mathbf{Dq_k})^H \mathbf{x}(n) \quad (5)$$

Based on (5), the MTSE is formulated as:

$$\hat{PSD}_{MTSE}(f_i) = \frac{1}{K}\sum_{k=0}^{K-1}|\kappa_k(f_i)|^2 \quad (6)$$

Indeed if there is only a vector $\mathbf{q_0}$ containing $1/\sqrt{N}$ 's as its elements, (6) becomes periodogram with rectangular window. And if $\mathbf{q_0}$ is a window function then (6) becomes the windowed periodogram. Hence, (6) can generally be considered as average of several periodograms with different windows. $\kappa_k(f_i)$ can be viewed as the output of $k^{th}$ bandpass filter of a group of bandpass filter banks with different filter response.

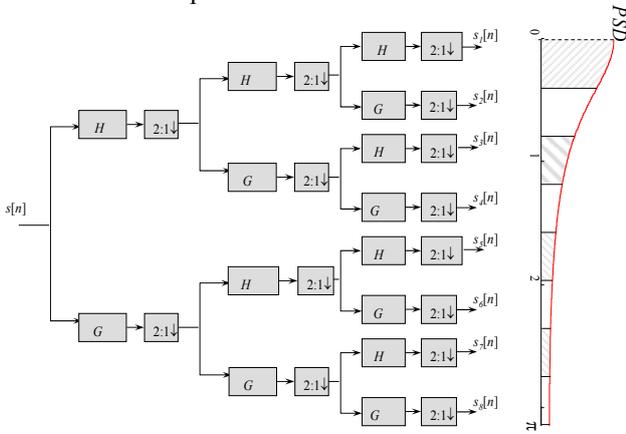

**Figure 1.** Wavelet packet decomposition of a signal. Here *H* and *G* denote the frequency responses of the low and high pass decomposition filters, respectively. The down arrows represent decimation by 2. The $s_i$'s denote the wavelet packet coefficients. Besides the decomposition, the Power Spectral Density (PSD) of the decomposed signal components in successive octave bands normalized to the Nyquist frequency is shown.

### III. WAVELET PACKET REPRESENTATION

The wavelet packet (WP) approach is a natural extension to the MTSE in the sense that this method also uses different orthogonal filters as prototype filters which are derived from tree structures constructed by cascading paraunitary filters. The algorithm of discrete wavelet packet transform and spectrum estimation is implemented by two-channel filter banks containing a half-band low and high pass filter duo [5]. The analysis of a signal is carried out by first decomposing the signal into an approximation containing coarse and detailed information. The coarse information corresponds to low pass filtering and the detailed information corresponds to high pass filtering. The scheme is then iterated successively on both the coarse and detailed versions. Through a hierarchical coding scheme, the signal to be encoded is successively split into high and low frequency components. The decomposition of the signal into different frequency bands with different resolutions is therefore obtained by successive high and low pass filtering of the signal. The number of successions is usually limited by the desired level of frequency resolution and available computational power. Figure 1 illustrates a level-3 decomposition procedure which generates eight wavelet packet coefficients.

#### A. Frequency Ordering of WP Coefficients

The frequency ordering of wavelet packet coefficients is not in a succession but rather in binary Gray code order. This is because the output of any 2-channel analysis is the result of low/high pass filtering followed by down sampling. Down sampling generates two new filter results with half the number of elements. In addition to this it also results in mirroring of the high pass components. This switches the order of low and high pass components in a subsequent decomposition. When the wavelet packet algorithm is recursively applied the frequency ordering of the resultant carriers follow the Gray code sequence [6]. This factor must be considered during the identification of used/unused bands.

#### B. Energy Conservation And Parseval Relation

In order to obtain valid estimates, the relationship between the signal amplitude and the WP coefficients needs to be defined. As already known, the Parseval relation proves that the Fourier transform is a lossless unitary transform. Likewise there is a need to assert if the wavelet packet transforms preserves energy too. In order to verify this we can start by representing a function $P(x)$ in Hilbert Space as linear combination of a basis function $\varphi_i(x)$:

$$P(x) = \sum_i \alpha_i \varphi_i(x) \quad (7)$$

It is clear from (7) that $\alpha_i$ can be obtained from inner product between basis function $\varphi_i(x)$ and function $P(x)$:

$$\alpha_i = \langle \varphi_i(x), P(x) \rangle \quad (8)$$

The norm of the function can also be computed from transform coefficients:

$$\|P(x)\| = \sum_i |\alpha_i|^2 = \sum_i |\langle \varphi_i(x), P(x) \rangle|^2 \quad (9)$$

By assuming that a function $Q(x)$ has transform coefficients $\beta_i$, we can derive the generalized Parseval equation by taking the inner product between two functions $P(x)$ and $Q(x)$ in Hilbert Space [7]:

$$\langle P(x), Q(x) \rangle = \sum_i \bar{\alpha}_i \beta_i = \sum_i \langle P(x), \varphi_i(x) \rangle \langle \varphi_i(x), Q(x) \rangle \quad (10)$$

In (10), $\bar{\alpha}_i$ indicates the complex conjugate of $\alpha_i$. In general, discrete wavelet transform pairs for discrete signal $x[n]$ can be represented as follows:

$$x[n] = \left(\sum_{j=1}^{J}\sum_{k=1}^{M/2^j} d_{j,k}\psi_{j,k}[n]\right) + \sum_{k=1}^{M/2^J} s_{J,k}\varphi_{J,k}[n] \quad (11)$$

$$d_{j,k} = \langle \psi_{j,k}, x \rangle, \; s_{J,k} = \langle \varphi_{J,k}, x \rangle \quad (12)$$

In (11) and (12), $J$ and $M$ are the decomposition level and total number of coefficients, respectively. $\psi_{j,k}$ and $\varphi_{J,k}$ denote the wavelet and scaling functions with $d_{j,k}$ and $s_{J,k}$ standing for their respective transform weights. Equations (11) and (12) are nothing but the synthesis and analysis equations, respectively. The first component of (11) is the detail part of signal $x[n]$, which is represented as linear combination of the wavelet function $\psi_{j,k}[n]$. And the second part of (11) is the coarse version of $x[n]$, which is represented as linear combination of scaling function $\varphi_{J,k}[n]$. If we have another signal, $y[n]$ with $d^{(i)}_{j,k}$ and $s^{(i)}_{J,k}$ as its WP coefficients, then the Parseval relation for $y[n]$ and $x[n]$ can be described using (10) as [7]:

$$\langle x, y \rangle = \sum_{n=-\infty}^{\infty} x[n]y[n] = \left( \sum_{j=1}^{J} \sum_{k=1}^{M/2^j} d^{(i)}_{j,k} d_{j,k} \right) + \left( \sum_{k=1}^{M/2^J} s^{(i)}_{J,k} s_{J,k} \right) \quad (13)$$

By taking $x[n] = y[n]$, the Parseval relation for the norm of $y[n]$ can be given as:

$$\|y\|^2 = \sum_{n=-\infty}^{\infty} |y[n]|^2 = \left( \sum_{j=1}^{J} \sum_{k=1}^{M/2^j} |d_{j,k}|^2 \right) + \sum_{k=1}^{M/2^J} |s_{J,k}|^2 \quad (14)$$

Equation (14) clearly illustrates the lossless nature of wavelet transform. Hence, the discrete wavelet transform preserves the time domain energy in wavelet domain. This lossless feature is really important and a fundamental reason why the spectrum sensing technique based on wavelet can be built.

### C. Spectral Energy from WP Coefficients

Since the Parseval relationship holds good for WP transform, the relationship between the energy ($E_{WPm}$) and power ($P_{WPm}$) for a WP node $m$ and total number of samples $N_{SAMPLES}$ can be described as:

$$P_{WPm}(watt) = \frac{E_{WPm}(Joule)}{N_{SAMPLES}} \quad (15)$$

The power spectrum density in the $m^{th}$ frequency band ($\hat{PSD}_{WPm}$) corresponding to $m^{th}$ WP node can be found by simply dividing $P_{WPm}$ by the frequency range ($f_{WP}$) spanned by single WP node as:

$$\hat{PSD}_{WPm}(watt/Hz) = \frac{P_{WPm}(watt)}{f_{WP}(Hz)} \quad (16)$$

## IV. RESULTS AND ANALYSIS

### A. Experiment Scenarios, Sources And Characteristics

To gauge the performance of the proposed WP technique, two types of sources are considered, namely, single tone and partial band. A single-tone source has all its energy at one frequency while the partial-band source has its energy spread over a continuous range of frequencies. In this work the single tone source is taken to be at normalized frequency $0.5\pi$ and the partial-band source in the normalized frequency band $[0.25\pi, 0.75\pi]$. To fine tune the estimation, different levels of WP decomposition are considered. The results of the experiments are compared with existing techniques such as Welch, Periodogram estimates, periodogram with windowing (Hann, Hamming and Blackman) and MTSE. Frequency selective wavelet, with regularity index of 12, filter length of 30 and transition band 0.1, is the filter bank of choice [8]. Regularity index is defined in terms of the number of times the wavelet is continuously differentiable and is a measure of smoothness of the wavelet [5]. Our reason to select frequency selective wavelet is because it offers a better performance than other wavelet families with considerably shorter filter length [9]. The number of samples in this experiment is 12800. In Welch approach, the input samples are divided into smaller segments and the periodogram for each segment is computed. The content of one segment overlaps with the content of neighboring segments. Later, the outputs of all periodograms are averaged. The overlap percentage and the length of each segment employed in Welch approach is 50% and 64 samples, respectively. This means the averaging approach performed in Welch is done over 399 realizations. Hamming window is used in this Welch approach.

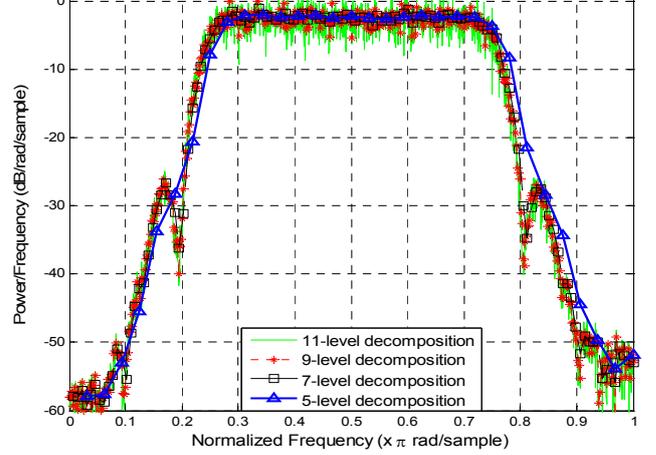

**Figure 2.** PSD estimates of partial band source according to various decomposition level of frequency selective wavelet based approach.

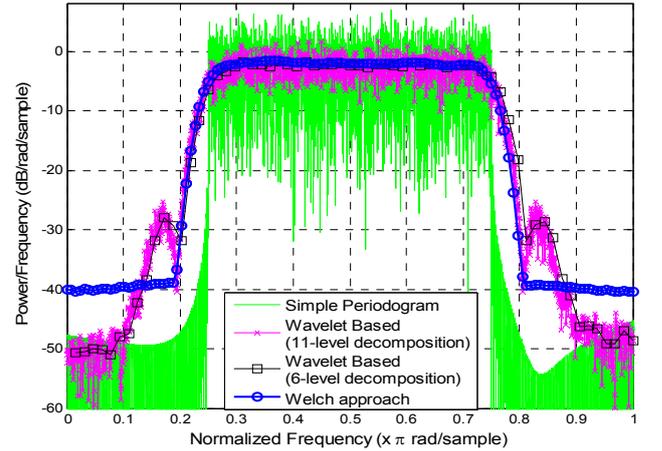

**Figure 3.** PSD estimates of partial band source according to two different decomposition level of frequency selective wavelet based approach, Welch approach and simple periodogram.

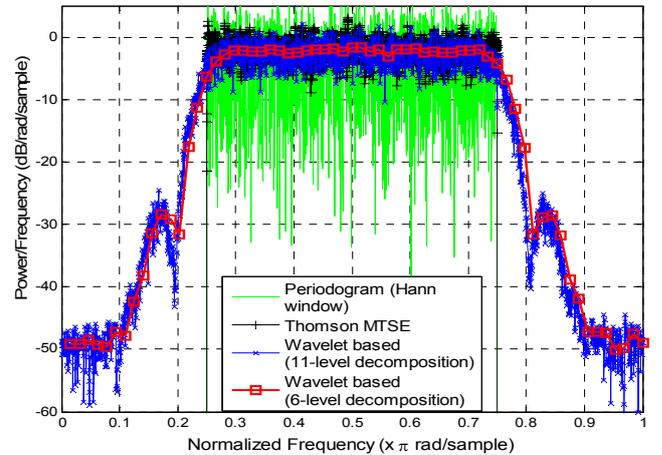

**Figure 4.** PSD estimates of partial band source according to two different decomposition level of frequency selective wavelet based approach, Thomson's MTSE and Periodogram using Hann window.

### B. Partial Band Source

Figures 2-5 provide the PSD estimate for partial band source with various techniques. The results are provided

in four different figures for ease of depiction. In figure 2, the wavelet based PSD estimates are displayed at 4 different decomposition levels while in the next three figures, only two decomposition levels are presented for the sake of clarity. The performance of the estimation techniques are evaluated with respect to four different metrics: side lobe suppression, variance of the estimated PSD in pass band and stop band, and transition between pass band and stop band (transition band). The best performing system is one which yields excellent transition band with good side lobe suppression and low stop/pass band variance. Indeed all these metrics may not be realized at the same instance and one may have to trade-off between the desirables to select the best system.

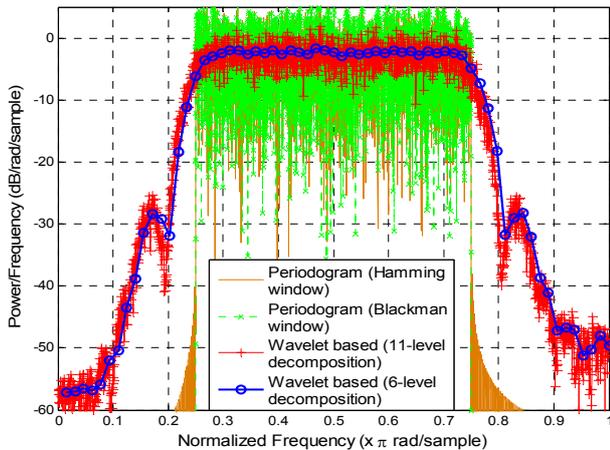

**Figure 5.** PSD estimates of partial band source according to different decomposition level of frequency selective wavelet based approach together with Periodogram using Hamming and Blackman window.

### B.1. Comparison with Welch and periodogram

The Periodogram has a good transition band but has a large variance in the passband. The Welch approach employs averaging and hence the variances are low, but then the transition band is poor. Of interest is the WP approach where one may increase or decrease the decomposition levels to achieve the desired variance of the estimated PSD. With a decrease in the depth of data decomposition the variance is significantly improved. A little price is however paid in terms of a decrease in the frequency resolution.

**Table 1.** Summary of comparison of wavelet packet spectrum estimation performance for estimation of partial band source with other techniques with regard to various performance metrics.

| Estimation Methods | Performance Measures | | | |
|---|---|---|---|---|
| | Side lobe Suppression | Variance in pass band | Transition Band | Variance in stop band |
| Welch | ≈ | - | ≈ | - |
| Periodogram | - | ++ | - | ++ |
| Periodogram with Window | - | ++ | - | ++ |
| MTSE | - | + | - | + |

### B.2. Comparison with Windowed periodogram and MTSE

Applying the window to the periodogram (Figure 4 and 5) reduces the side lobes in the estimates but it does not solve the large variance in the estimates. In fact all of the windows introduced here have variances much larger than 11-level WP estimation scheme. Lastly, the MTSE estimates have good frequency resolution but they too suffer from significant estimate variance.

Table 1 summarizes the comparison of the wavelet packet approach with other approaches. The notations +, - and ≈ indicate whether the WP approach performs favorably, negatively or similar in comparison to the other method. It is clear from the PSD curves and the table that the WP approach performs quite reasonably with all the metrics defined and its operation compares favorably with most of the existing approaches.

### C. Single Tone Source

With regard to the estimation of single tone source, the performance metrics used are: mean power in stop band, variance in stop band, frequency resolution and side lobe suppression. Figure 6-9 show the PSD estimates for single tone source.

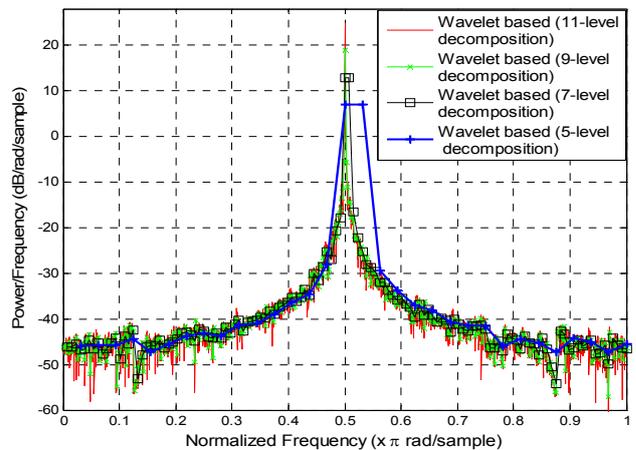

**Figure 6.** PSD estimates of single tone source according to various decomposition level of frequency selective wavelet based approach.

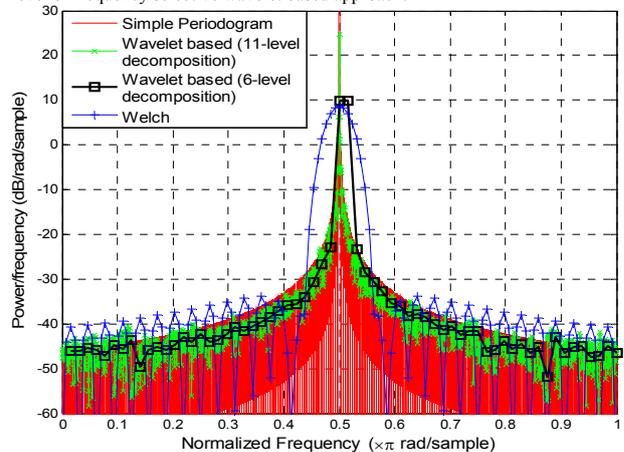

**Figure 7.** PSD estimates of single tone source according to two different decomposition level of frequency selective wavelet based approach, Welch approach and simple periodogram.

### C1. Comparison with Welch and periodogram

The result of this experiment exemplifies the fact that the wavelet based estimates have characteristics in between that of periodogram (excellent frequency resolution but large stop band variance) and Welch approaches (low stop band variance but poor frequency resolution). The Welch approach windows and averages and hence tends to *smear* the details that one might want to see in the PSD. Hence, its frequency resolution is poor. The alternative, if periodogram is chosen, is to not be sure if features one sees in the PSD are real or merely spurious noise, so one is basically stuck between a rock and a hard place. In this regard one can say that the performances of the WP approach can be operated between the strengths of Welch approach (low variance) and that of periodogram (excellent resolution) without compromising too much on either of these metrics by merely increasing/decreasing the levels of decomposition.

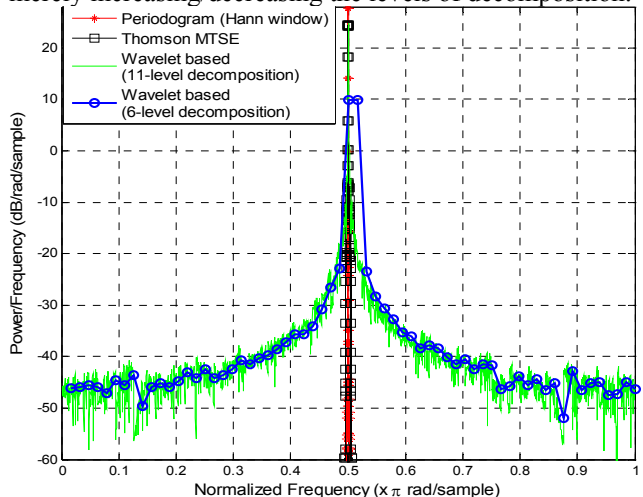

**Figure 8.** PSD estimates of single tone source according to different decomposition level of frequency selective wavelet based approach, Thomson's MTSE and Periodogram using Hann window.

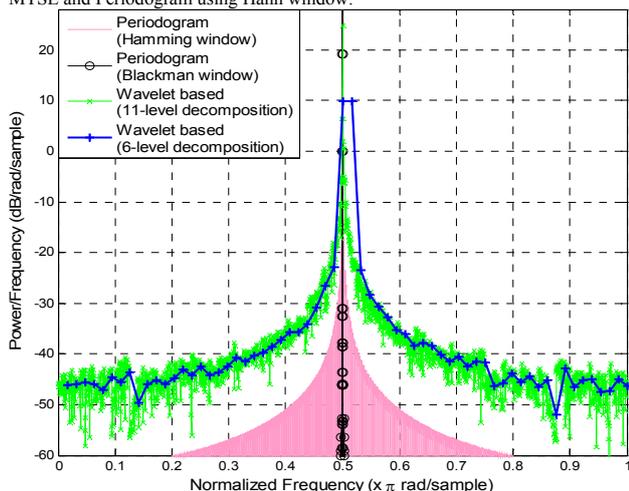

**Figure 9.** PSD estimates of single tone source according to different decomposition level of frequency selective wavelet based approach, Periodogram using Hamming and Blackman window.

In this single tone source estimation, higher WP decomposition level leads to better frequency resolution. However, as the decomposition level is increased from 5 to 11, the variance of the estimate increases as well. The wavelet based estimates tend to approach periodogram estimate for higher order decomposition levels and approaches the Welch based estimate for lower orders. With regard to suppression of sidelobes, the wavelet based estimates achieve slightly greater suppression than the Welch method.

*C2. Comparison with Windowed periodogram and MTSE*

Figure 8 and 9 show the impact of windowing on the reduction of side lobe level of the estimates. Periodogram based estimate using Hann, Hamming and Blackman window seems to offer better frequency resolution than wavelet based estimate. However, it also appears that the frequency resolution of wavelet based estimate tends to approach that of those three periodogram estimates as the WP decomposition level is increased. This means that there is a promise offered by wavelet based estimates as long as the required decomposition level can be fulfilled. The MTSE also generally offers a better frequency resolution and much better side lobe rejection.

Table 2 summarizes the comparison of the wavelet packet approach with other approaches for estimation of single tone source.

**Table 2.** Summary of comparison of wavelet packet spectrum estimation performance for estimation of single tone source with other techniques with regard to various performance metrics.

| Estimation Methods | Performance Measures | | | |
|---|---|---|---|---|
| | Mean Power in Stop band | Variance in stop band | Frequency Resolution | Side lobe Suppression |
| Welch | + | ≈ | ++ | + |
| Periodogram | ≈ | ++ | ≈ | ≈ |
| Periodogram with Window | - | ++ | ≈ | - |
| MTSE | - | + | ≈ | - |

## V. CONCLUSION

In this article, the application of wavelet packet transform for spectrum estimation was proposed and investigated. First traditional spectrum estimation techniques, such as Periodogram, Blackman-Tukey, Welch and Multi taper estimation, were stated as filter bank analysis problem. Then the Wavelet Packet based spectrum estimation, implemented using paraunitary filters, and was presented as advancement to the existing approaches. To gauge the performance of the proposed technique, two test sources with single tone and partial band characteristics were chosen. And the performance metrics used were variance and resolution of the estimate and side-lobe suppression. The studies showed that the proposed estimator operated well for all types of sources and its performances were comparable or at times even better than other existing approaches.